\documentclass[authoryear, 12pt]{elsarticle}
\usepackage[english]{babel}
\usepackage{amsmath}
\usepackage[toc,page]{appendix}
\usepackage{amssymb}        
\usepackage{amsthm}        
\usepackage{natbib}         
\usepackage{mathtools}
\usepackage{graphicx}       
\usepackage{subfigure}      
\usepackage{color}
\usepackage{gensymb} 
\journal{Journal of the Mechanics and Physics of Solids, accepted for publication}

\title{Strengthening of Al-Cu alloys by Guinier-Preston zones: predictions from atomistic simulations} 

\author{G. Esteban-Manzanares$^{1, 2}$}
\author{B. Bell{\'o}n$^{1, 2}$}
\author{E. Mart{\'\i}nez$^{3}$}
\author{I. Papadimitriou$^{1}$}
\author{J. LLorca$^{1, 2, }$\corref{cor1}}
\address{$^1$ IMDEA Materials Institute, C/ Eric Kandel 2, 28906, Getafe, Madrid, Spain. \\  $^2$ Department of Materials Science, Polytechnic University of Madrid/Universidad Polit\'ecnica de Madrid, E. T. S. de Ingenieros de Caminos. 28040 - Madrid, Spain. \\  $^3$ Theoretical division, T-1, Los Alamos National Laboratory, Los Alamos 87545 NM, USA.}

\cortext[cor1]{Corresponding author}

\begin{document} 

\begin{abstract}

A  scale bridging strategy  based in molecular statics and molecular dynamics simulations in combination with transition state theory has been developed to determine the flow stress of Al-Cu alloy containing Guinier-Preston zones. The athermal contribution to the flow stress was determined from the Taylor model, while the thermal contribution was obtained from the obstacle strength and the free energy barrier. These two magnitudes were obtained by means of molecular statics and molecular dynamics simulations of the interaction of edge dislocations with Guinier-Preston zones in two different orientations. The predictions of the model were compared with experimental data and were in reasonable agreement, showing the potential of atomistic simulations in combination with transition state theory to predict the flow stress of metallic alloys strengthened with precipitates.
\end{abstract}


\maketitle

\newpage
\section{Introduction} \label{Intro}

Precipitation hardening is one of the most efficient strategies to increase the yield strength of metallic alloys \citep{A85,N97}. Dislocation glide along the crystallographic planes and directions which require the lowest critical resolved shear stress is hindered by the presence of nm-sized precipitates and the overall strength of the alloy is increased. Large precipitates ($>$ 50 nm) with incoherent interfaces cannot be sheared by the dislocations and they are overcome by the formation of Orowan loops \citep{O34}. This process can be modelled within a continuum framework from the elastic interaction of the dislocation line with the precipitate, leading to simple analytical models to determine  the yield strength as a function of the spacing, size and shape of precipitates \citep{KN63, BKS73, N12}. These analytical models were able to rationalize the main experimental trends of the effect of the precipitate size and spatial distribution on the yield strength but could not give quantitative estimations. They have been obtained more recently through dislocation dynamics simulations which can include the influence of the elastic mismatch, precipitate morphology (e.g. shape and orientation) as well as of the stress-free transformations strains associated with the formation of the precipitates \citep{XSC04, QB09,  MND11, GFM15, ZSD17, SEP18}.

Metallic alloys are also strengthened by small precipitates ($<$ 20 nm) that can be sheared by dislocations. This is the case of Guinier-Preston (GP) zones in many metallic alloys \citep{BFK61, RBP18}, $\theta''$ and $\Omega$ precipitates in Al-Cu and Al-Cu-Mg alloys \citep{N14}, $\beta'_1$-MgZn$_2$ in Mg-Zn alloys \citep{WS15}, $\beta_1$ in Mg-Nd alloys \citep{ZWZ18, HYQ19}, $\gamma'$-Ni$_{3}$(Al, Ti)  and $\gamma''$-Ni$_3$Nb in Ni-based superalloys. \citep{XCC08, CGJ15}, etc. The different factors that control the CRSS to shear the precipitate in this case (chemical, stacking-fault, modulus, coherence and order strengthening)  have been evaluated using continuum models based on the dislocation line tension \citep{A85, N97} or, more recently, through dislocation dynamics simulations \citep{VDR09, GFM15, HZT12, HRU17}. These strategies were also able to capture the experimental trends in the case of Ni-based superalloys, although direct comparisons were more difficult because some of the critical parameters for these models (stacking fault energy of the precipitate, coherence strains, antiphase boundary energy) are not always easy to estimate. Moreover, the validity of these approaches can be questioned when the continuum hypothesis is no longer applicable due to the small dimensions of the precipitate. The paradigmatic case are the GP zones formed by a single layer of atoms coherent with the matrix and with diameter of 2 to 4 nm \citep{RBP18}. Shearing of the precipitate is controlled by the atomic interactions between matrix and solute atoms, leading to a very complex energy landscape that cannot be captured by simple continuum line tension models.

Atomistic simulations using molecular statics (MS) and dynamics (MD) seem to be the appropriate tool to simulate the interaction of dislocations with GP zones or very small precipitates \citep{HSC00, SW10, BTM11, LLH14, yanilkin2014dynamics, LGL16, EMS19} and nm-sized voids \citep{monnet2010mesoscale} because they can capture the interatomic interactions. Nevertheless, these approaches provide results that are valid at 0K (MS) or at extremely high strain rates for finite temperatures (MD). This information is useful to understand the interactions mechanisms in these two extreme conditions but cannot be directly extrapolated to provide quantitative values of the influence of the precipitates on the CRSS as a function of strain rate and temperature under standard loading conditions. Thus, new strategies are needed to upscale the information obtained from atomistic simulations to the mesoscale. 

Dislocation/precipitate interaction is a thermally-activated process \citep{K75} and  the stress necessary to overcome the precipitate as a function of the strain rate and temperature is given by the free energy barrier, $\Delta G$. According to the transition state theory \citep{GLE41},  the  energy barrier is only a function of the applied stress and becomes 0 at a given stress named the obstacle strength, $\tau_0$. Thus, the goal of atomistic simulations is to determine $\tau_0$ in the absence of thermal energy and the variation of $\Delta G$ with the applied stress $\tau$. The former can be achieved by means of MS simulations \citep{TBM08, monnet2015multiscale, SW10, EMP19} while different strategies have been proposed for the latter. For instance, \cite{saroukhani2016harnessing} explored the application of the harmonic transition state theory to determine the energy barrier of GP zones in an Al-Cu alloy. Under this assumption, the free energy barrier can be obtained from the activation potential energy (that can be computed using minimum energy path techniques, such as the nudged elastic band method) while the entropic activation only depends on the vibrational entropy and anharmonic effects are neglected. However, the predictions based on these assumptions were grossly inaccurate, as compared with direct MD simulations, and the differences were attributed to the presence of anharmonic effects. Better results were obtained through the application of interface sampling techniques, which are not based on the minimum energy path calculations. \cite{monnet2010mesoscale} obtained the influence of the shear stress on the free energy barrier by means of direct MD simulations in the case of voids sheared by dislocations. This approach is computational costly because many simulations are necessary for each combination of applied stress and temperature to have statistically significant results of the rate at which the dislocations overcome the obstacle.

\cite{EMS19} analyzed recently the interaction of dislocations with GP zones in Al-Cu by means of MS and MD. They found that the energy landscape for these interactions was very rough and standard MS simulations (based on the conjugate gradient algorithm) were not always able to find the minimum energy path because they were often trapped in local minima. As a result, the interaction mechanisms were unrealistic and the obstacle strength was overestimated. These limitations were overcome through the combination of MS with thermal annealing to overcome the local minima. Under these conditions, it was shown that the variation of free energy with the strain obtained from direct MD simulations was equivalent to the predictions provided by harmonic transition state theory from activation potential energy given by minimum energy path techniques for one particular GP zone orientation. Following this previous work, the interaction mechanisms between dislocations and GP zones with different orientations were analyzed in detail in this investigation and the corresponding values of the obstacle strength and of the energy barriers were obtained. They were used to develop of continuum model of the flow stress of Al-Cu alloys reinforced with GP zones that included the influence of athermal and thermal contributions, the latter using the hypothesis of the transition state theory \citep{K75}. Finally, the model results were compared with those obtained in an Al-4 wt.\% Cu alloy aged at room temperature, which contained a homogeneous distribution of GP zones of $\approx$ 4 nm in diameter \citep{RBP18} and whose yield strength has been measured as a function of temperature \citep{BFK61}. The paper is organized as follows. The computational methodology is presented in section \ref{Method} and the mechanisms of dislocation/GP zone interaction for both orienations variants are presented and compared in section \ref{Result}, together with  the results for the obstacle strength and free energy. This information is used in section \ref{FlowModel} to develop  a model to predict the CRSS in Al-Cu alloys containing GP zones. Finally, the conclusions are summarized in section \ref{sec6:Conclusion}.

\section{Atomistic model and simulation methodology} \label{Method}

\begin{figure}[!]
\centering
\includegraphics[width=0.85\textwidth]{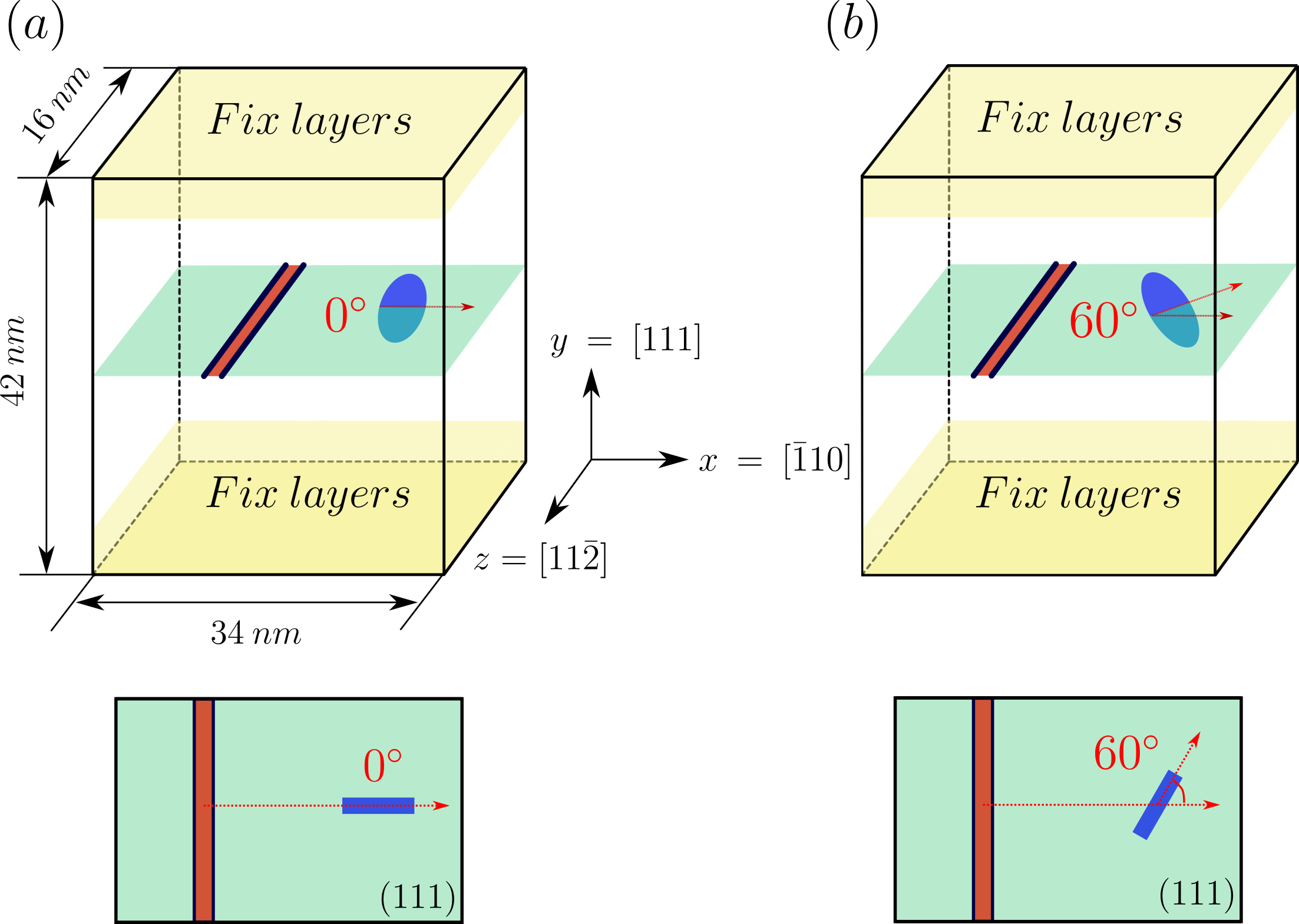} 
\caption{(a) Schematic of the atomistic simulation domain for the orientation GP0. (b) Schematic of the atomistic simulation domain for the orientation GP60.}
\label{configuration}
\end{figure}

GP zones develop in Al-Cu alloys after ageing at ambient temperature. They are single-layer disks of Cu atoms that grow parallel to the \{100\} planes of the cubic Al FCC lattice with an average diameter of the order of 4 nm  \citep{N14, RBP18}. All the atomistic simulations were carried out in a parallelepipedic domain of volume of 34 $\times$ 42 $\times$ 16 nm$^{3}$ (Fig. \ref{configuration}). These dimensions were chosen following the results reported by \cite{SC15} to minimize the image stresses in the simulation box. The $x$, $y$ and $z$ axes of the domain were parallel to the $[\bar{1}10]$,  $[111]$  and $[11\bar{2}]$ orientations of the FCC lattice.  The simulation domain contained an edge dislocation in the (111) glide plane (dissociated in leading and trailing Shockley partials) and a GP zone. Periodic boundary conditions were applied along the $x$ and $z$ axes, so the simulations stand for a periodic array of dislocations and precipitates \citep{osetsky2003atomicI}. Two different GP zone orientations were considered. In the first one (denominated GP0), the Burgers vector of the edge dislocation is parallel to the intersection of the GP zone in the slip plane $0^{\circ}$, Fig. \ref{configuration}(a). This configuration stands for the GP zones parallel to the (001) plane. In the second orientation (denominated GP60), the Burgers vector and the intersection of GP zone with the slip plane form an angle of 60$^{\circ}$, Fig. \ref{configuration}(b). This configuration includes the GP zones parallel to either (010) or (100) planes. The analysis of the dislocation/precipitate interaction in the GP60 orientation was already reported \citep{EMS19} and the atomistic simulations will be mainly focussed in the GP0 orientation. Nevertheless, the simulation results for GP0 will be compared with those obtained with GP60 and both are necessary to predict the flow stress.

The edge dislocation was introduced in the domain by applying the perfect dislocation displacement field for an isotropic elastic medium with ATOMSK \citep{hirel2015atomsk}. After the dislocation was introduced, the system was optimized using the conjugate gradient algorithm at constant volume, followed by another relaxation at zero Virial stresses. This led to the dissociation of the edge dislocation into two Shockley partials. The GP zone of 4 nm in diameter was  introduced by substituting the corresponding layer of Al atoms by Cu atoms and the energy of the system was minimized again, leading to a mechanical equilibrium configuration between the dislocation and the GP zone. Deformation was introduced by applying an atomic displacement along the $x$ axis and parallel to the slip plane to five layers of atoms on the top surface of the domain, while the five layers of atoms at the bottom surface remained fixed during the simulations. 

For each orientation, two different atomistic simulations were carried out. The obstacle strength was determined by means of MS simulations combined with thermal annealing, following the strategy developed in \cite{EMS19}. Classical MD simulations were used to establish the rate at which dislocations overcome the precipitate as a function of the applied strain and temperature. From this information,  the activation free energy was obtained within the framework of transition state theory. All atomistic simulations  were carried out using LAMMPS \citep{plimpton2007lammps}. OVITO was utilized to evaluate the results \citep{stukowski2009visualization}. The angular dependent potential (ADP) for the Al-Cu system developed by \cite{apostol2011interatomic} was employed in all the calculations.

\subsection{Determination of the obstacle strength} \label{GP0Ener}

From the initial configuration,  a shear strain was imposed to the simulation domain by applying a displacement of 0.25 $\mathrm{\AA}$ to the upper layer of atoms. Afterwards, the energy of the system was minimized using the conjugate gradient algorithm and the energy landscape was explored through thermal annealing. To this end,  a  linear temperature ramp from 100 K to 500 K was applied in 0.4 ns using a Langevin thermostat within the canonical ensemble (NVT). After the ramp, the temperature was held constant at 500 K for 2 ns and the per-atom coordinates  were stored each 0.02 ns. The energy of all the configurations stored during thermal annealing was optimized and the configuration with the minimum energy was selected for the next step, which began with the application of the shear displacement to the atoms in the upper layer. This process was repeated until the dislocation overcome the GP zone and the maximum shear stress attained during the process was the obstacle strength, $\tau_0$. The annealing temperature and time were chosen to ensure that they did not bias the sampling procedure \citep{EMS19}. Simulations without thermal annealing were also carried out for comparison.

\subsection{Determination of the rate}

Transition state theory establishes that the rate $\Gamma$ at which an ergodic system crosses a first order dividing surface is given by 

\begin{equation}
\Gamma = \nu \exp \left({\frac{-\Delta G(\tau)}{k_bT}}\right)
\label{TST-G}
\end{equation}

\noindent where  $G$ is the Gibbs free energy barrier (which depends on the applied stress $\tau$), $\nu$ the attempt frequency, $k_b$  the Boltzmann constant, and $T$ the absolute temperature. Within this scenario, MD simulations were used to obtain the rate at which edge dislocations overcome the GP zones as a function of the applied strain and temperature. It should be noted that if the MD simulations were carried out at constant strain $\gamma$, eq. \eqref{TST-G} can be expressed as

\begin{equation}
\Gamma =\nu\: \exp{\left(\frac{-\Delta F(\gamma)}{k_b T}\right)}
\label{TST-F}
\end{equation}

\noindent where $\Delta F(\gamma)$ is the Helmholtz activation free energy. $\Delta F(\gamma)$ is the energy barrier due to the presence of the obstacle in absence of external work and is given by
\begin{equation}
\Delta F(\gamma) =  \Delta U (\gamma) - T \Delta S_\gamma (\gamma, T)
\label{DeltaF}
\end{equation}

\noindent where $\Delta U (\gamma)$ stands for the activation energy, which depends on the applied strain, and $\Delta S_\gamma(\gamma, T)$ is the activation entropy at constant strain, which may depend on the applied strain and temperature. \cite{SR11} demonstrated that that $\Delta G(\tau) \approx \Delta F(\gamma)$  when the volume of the crystal is much larger than the activation volume of the thermally activated process and $\tau$ and $\gamma$ are conjugate variables. Under these conditions, the rate at which the obstacle is overcome is independent of whether the crystal is subjected to a constant stress or to a constant strain that corresponds to the same stress. Thus,  MD simulations at constant shear strain were carried out within the NVT ensemble at 400, 450, 500, 550 and 600 K to determine the energy barrier.  The time needed for the dislocation to overcome the precipitate, $t^s$, was obtained for each case. Eight non-correlated simulations were carried out for each strain and temperature to account for the statistics of the process. Hence, the rate, $\overline \Gamma$, was determined by $8/\sum t^s_i$.  It should be noted that hydrostatic stresses developed during the MD simulations within the NVT ensemble due to thermal expansion effects. Nevertheless, they do not influence the rate at which dislocations overcome the obstacle because dislocation slip is controlled by the shear stresses. This hypothesis was verified by carrying out  MD simulations using a slightly different strategy in selected cases. The temperature was initially increased from 0K up to the simulation temperature using a NPT ensemble at zero pressure. Thus, the volume changed to reach the simulation temperature with negligible pressure. Then, the NVT ensemble was used to perform the MD simulation at a given value of the applied strain to determine the rate. The rates obtained with both strategies were equivalent .

\section{Results and discussion} \label{Result}

\subsection{Obstacle strength and dislocation/GP interaction}\label{MS-MSTA}

\begin{figure}[t]
\centering
\includegraphics[scale=0.8]{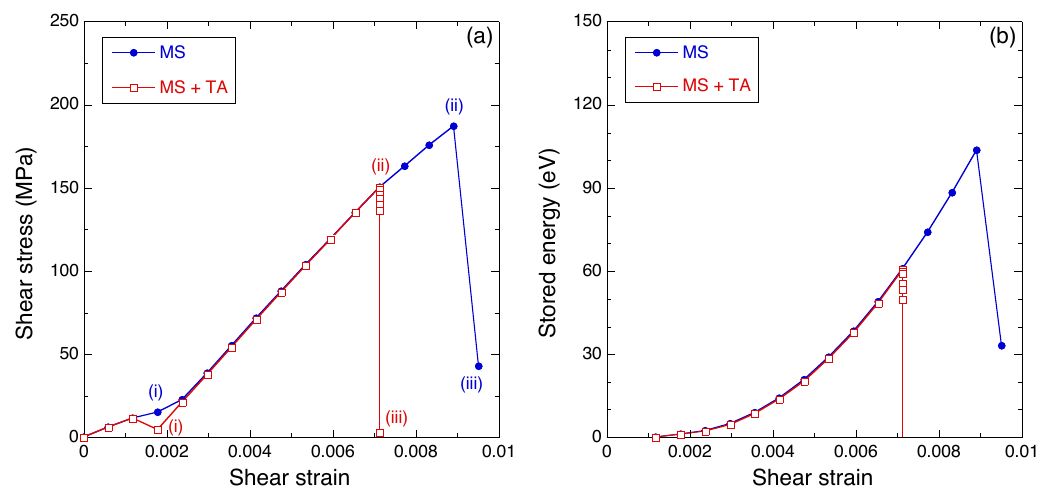} 
\caption{(a) Shear stress {\it vs.} shear strain curves corresponding to the GP0 orientation. (b) Energy stored {\it vs.} shear strain curves corresponding to the GP0 orientation. Results from MS simulations and MS simulations combined with thermal annealing (MS + TA) are plotted for comparison.}
\label{MSTA_GP0}
\end{figure}

\begin{figure}[t]
\centering
\includegraphics[width=\textwidth]{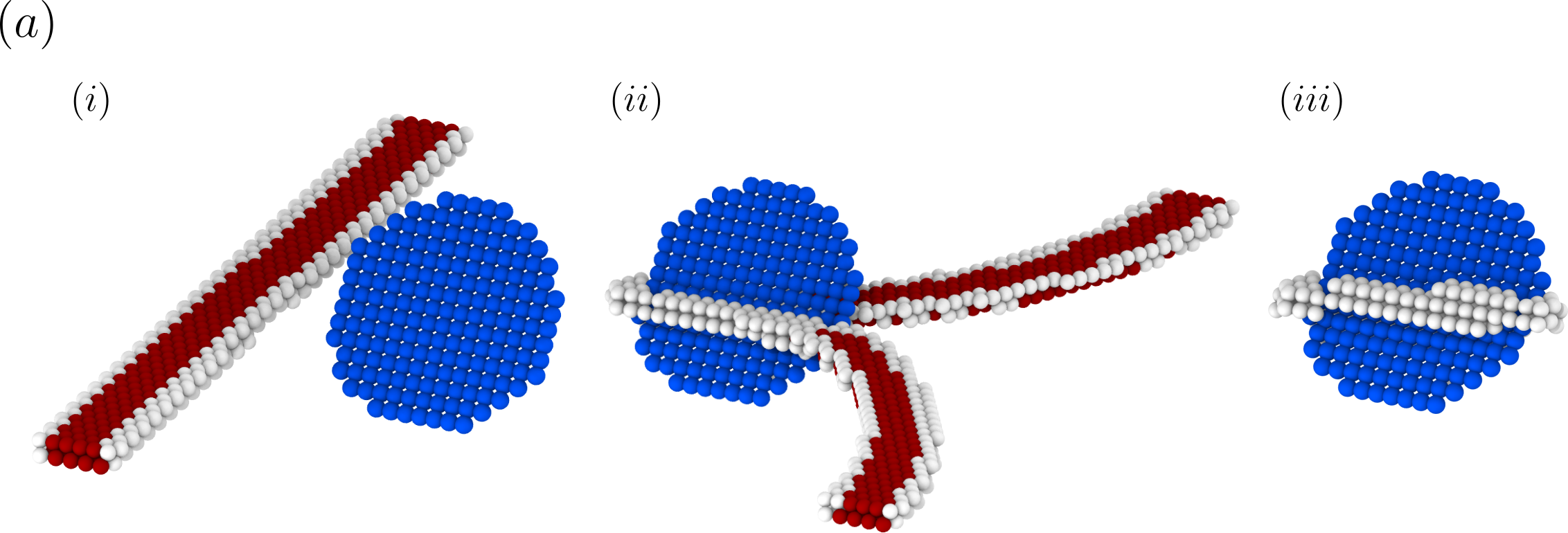} 
\includegraphics[width=\textwidth]{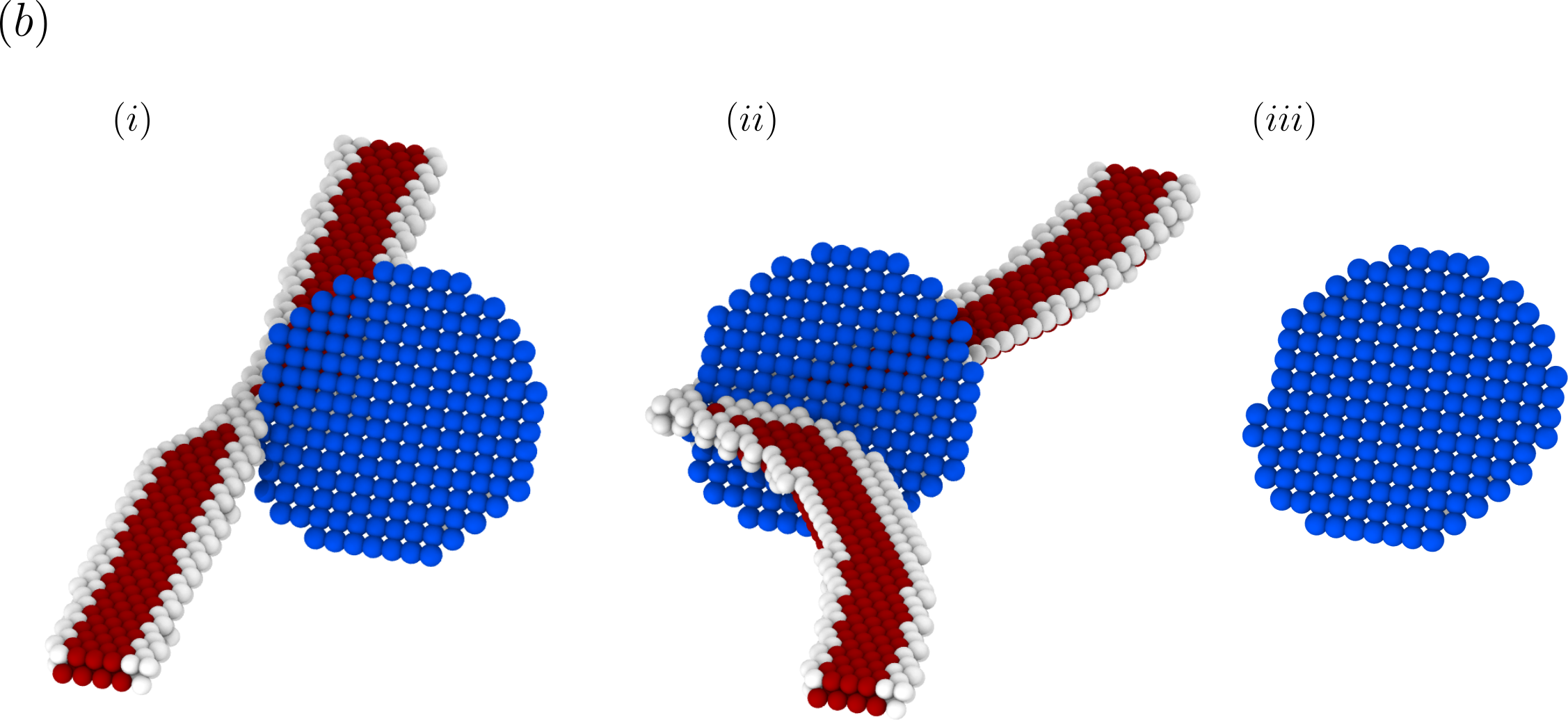} 
\caption{Atomistic representation of the dislocation-GP zone successive interactions in the GP0 orientation.  (a) Molecular statics. $(b)$ Molecular statics in combination with thermal annealing. The plots $(i)$, $(ii)$ and $(iii)$ correspond to the points marked with the same letters in Fig \ref{MSTA_GP0}.}
\label{MS_atom}
\end{figure}

The dislocation/GP zone interactions were analyzed by means of MS simulations with and without thermal annealing. The shear stress-strain curves and the energy stored-strain curves obtained from the simulations are plotted in Figs. \ref{MSTA_GP0}(a) and (b), respectively, for the GP0 configuration. Three different regions can be observed along the MS and MS + TA stress-strain curves in Fig. \ref{MSTA_GP0}(a). Both MS and MS + TA curves are initially superposed, as the dislocation starts to move towards the GP zone. The initial interaction between the dislocation and the GP zone was noticed when the applied shear strain $\gamma$ reached 0.18\%  (marked with $i$) and the simulations carried out with thermal annealing showed that the dislocation was attracted by the GP zone, leading to a reduction in the shear stress. This phenomenon was not captured, however, in the absence of thermal annealing. The corresponding atomistic representation is illustrated in Figs. \ref{MS_atom}$(a)$ and $(b)$ for MS and MS + TA simulations, respectively. The dislocation is not in contact with the GP zone in the case of MS (Fig. \ref{MS_atom}(a) $(i)$), while the dislocation is in contact with the GP zone in the case of MS + TA (Fig. \ref{MS_atom}(b) $(i)$). This situation with lower energy was attained by overcoming small barriers in the vicinity of the initial point obtained by regular MS. Although the differences between both configurations are very small from the viewpoint of the total stored energy in Fig. \ref{MSTA_GP0}(b), they mark the beginning of the divergence between both paths.

 Afterwards, the elastic interaction of the dislocation with the GP zone is shown in the hardening of the shear stress-strain curves. In the MS simulation, the dislocation bows around the GP zone, leading to the formation of an Orowan loop around the GP zone after the dislocation has overcome the obstacle, see $(ii)$ and $(iii)$ in Fig. \ref{MS_atom}(a). However, the GP zone is sheared by the dislocation in the case of the MS + TA simulation, Fig. \ref{MS_atom}(b) $(ii)$ and $(iii)$. The obstacle strength, $\tau_0$, in this case is only 150 MPa, significantly lower than the one predicted for the formation of an Orowan loop in the MS simulation. It is also interesting to notice that the shearing of the GP zone in the MS + TA simulations takes place at constant strain and that both stress and energy go to 0 once they GP zone has been sheared. Shearing at constant strain while the applied stress decreases indicates that the largest energy barrier is found when the trailing dislocation enters the GP zone.  Once this barrier is overcome, the dislocation progresses and the shear stress drops to zero while the structure of the GP zone shows the displacement by one Burgers vector in the glide plane, Fig. \ref{MS_atom}(b) $(iii)$. This final stage is different from the one obtained in the MS simulations, in which the stress does not drop to zero after the dislocation overcomes the GP zone because of the presence of the Orowan loop. 
 
\begin{figure}[t]
\centering
\includegraphics[scale=0.8]{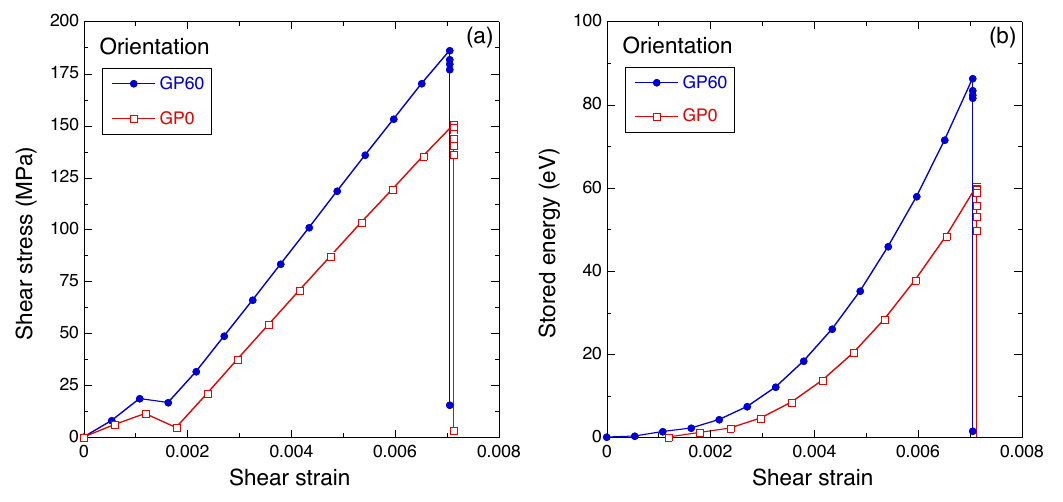} 
\caption{(a) Shear stress {\it vs.} shear strain curves for GP60 and GP0 orientations. (b) Stored energy {\it vs.}  shear strain curves for for GP60 and GP0 orientations. All the results were obtained with  MS+TA simulations.}
\label{GP_ss}
\end{figure}

These results are similar to those reported previously for the GP60 orientation \citep{EMS19}: shearing of the GP zone by the dislocation was only observed in the MS + TA simulations, and the obstacle strength was also significantly reduced as compared with the MS simulations. For the sake of completion, the shear stress-strain curves and the energy stored-strain curves obtained in the MS + TA simulations are plotted in Figs. \ref{GP_ss}(a) and (b), respectively,  for GP0 and GP60 orientations. Both curves show the same features for both GP zone orientations. The dislocation is initially attracted to the GP zone and further deformation leads to hardening as the dislocation tries to overcome the obstacle. This is achieved by shearing, which takes place a constant strain and leads to a relaxed system in which the GP zone has been sheared by a Burgers vector. The obstacle strengths, $\tau_0$, corresponding to the maximum in the shear stress are indicated in Table \ref{Param}. It should be noted that the obstacle strength and the energy stored before shearing in the GP0 orientation are lower than those in the GP60 orientation. The obstacle strength is lower in the GP0 orientation because the length of the dislocation line that interacts with the GP zone is shorter in the GP0 orientation (the GP zone is perpendicular to the dislocation line, Fig. \ref{configuration}(b) than in the GP60 orientation, where it is proportional to the GP zone diameter. This leads to lower initial stresses in the GP0 orientation when the dislocation approaches the GP zone and also to a lower elastic hardening as the dislocation tries to overcome the GP zone.

\begin{figure}[!]
\centering
\includegraphics[width=\textwidth]{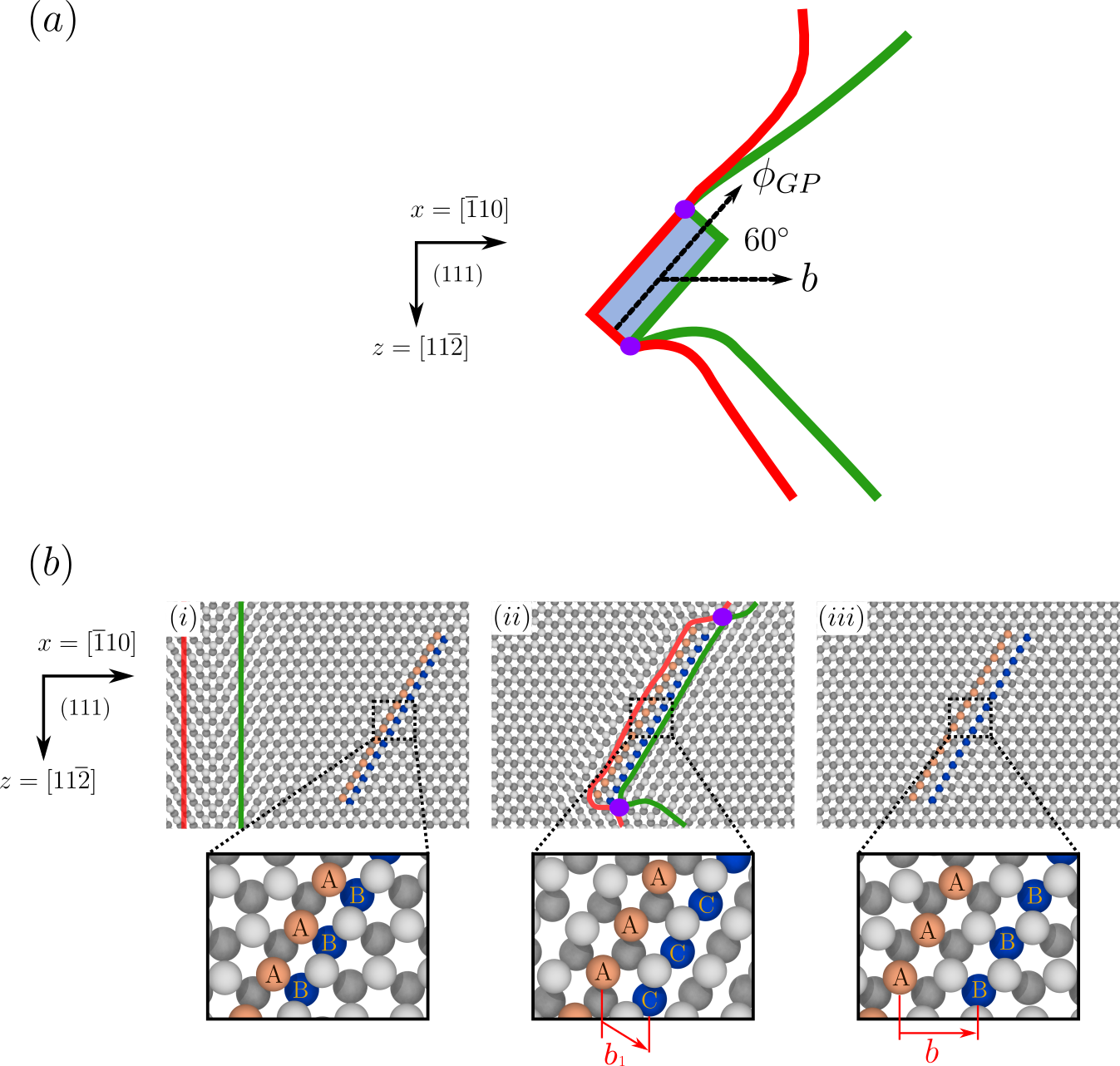} 
\caption{(a) Schematic representation of the interaction between Shockley partial dislocations with the GP zone in the GP60 orientation. (b) Atomistic representation of the shearing of the GP zone by the Shockley partial dislocations along the (111) plane. $(i)$ stands for the initial configuration before the Shockley partials have reached the GP zone. $(ii)$ shows the atomic arrangement when the leading partial has sheared the GP zone while $(iii)$ depicts the atom positions when the trailing partial has sheared the GP zone. The leading partial is shown as a green line and the trailing partial as a red line. Blue dots stand for Stroh nodes. Al atoms are shown as grey spheres. Cu atoms that are not displaced during shearing are shown as brown spheres while Cu atoms that are displaced during shearing are shown as blue spheres. These atoms move by $b_1$ from $(i)$ to $(ii)$ and by $b$ from $(ii)$ to $(iii)$, as shown by the red arrows. Atoms with light colors are in the upper plane and with dark color in the lower plane.} 
\label{GP60}
\end{figure}

There are also differences in the shearing mechanisms between both orientations. In the case of the GP60 orientation, the leading partial has to overcome a small local maximum in the stress when the applied strain was 0.1\% (Fig. \ref{GP_ss}(a). Afterwards, the leading Shockley partial sheared the GP zone, reaching a metastable position denoted by the minimum in the shear stress-strain curve when the applied shear strain was 0.16\%. The current configuration of the Shockley partials at this position is schematically depicted in Fig. \ref{GP60}(a). At this stage, the leading Shockley partial  (green line) has already sheared the GP zone while the trailing partial (red line) is stopped at the interface. Both partial dislocations join together to form two Stroh nodes (blue dots) but they remain dissociated within the GP zone. The displacement of the Cu atoms  in the (111) plane during the shearing of the GP zone by the leading partial is detailed in Fig. \ref{GP60}(b) from $(i)$ to $(ii)$. Cu atoms that do not move during shearing are marked as brown spheres while those that are displaced are blue spheres. Shearing by the leading partial led to the displacement of the Cu atoms in blue by $b_1$ from the $B$ position to the $C$ position in the FCC lattice, where $b_1 = \frac{a}{2}[11\bar2]$ is the Burgers vector of the partial dislocation and $a$ the lattice parameter. The atomic displacements during the shearing of the GP zone by the trailing partial are depicted in Fig. \ref{GP60}(b) from $(ii)$ to $(iii)$. The blue Cu atoms in $B$ positions move to $C$ positions, leading to the full shearing of the GP zone by a Burgers vector $b$. It should be noted that the highest energetic barrier to shear the precipitate in this configuration is found to introduce the trailing partial in the precipitate. Moreover, the detailed analysis of the atomistic simulations showed that shearing of the GP by the leading partial took place by the simultaneous movement of all the Cu atoms from $B$ to $C$ while the displacement from $C$ to $B$ follows a staircase sequence in the case of the trailing partial. Thus, the analysis of the atomic sequence of the shearing of GP zone by dislocations in the GP60 orientation shows that there is a major energetic barrier (corresponding to the penetration of the trailing partial in the GP zone) followed a finite number of very small energetic barriers as the trailing partial shears the precipitate. Nevertheless, the magnitude of these barriers is very small compared with the large barrier to introduce the trailing partial into the GP zone and the whole process could be modelled with only one energetic barrier within the framework of transition state theory \citep{EMS19}.

\begin{figure}[!]
\centering
\includegraphics[width=0.75\textwidth]{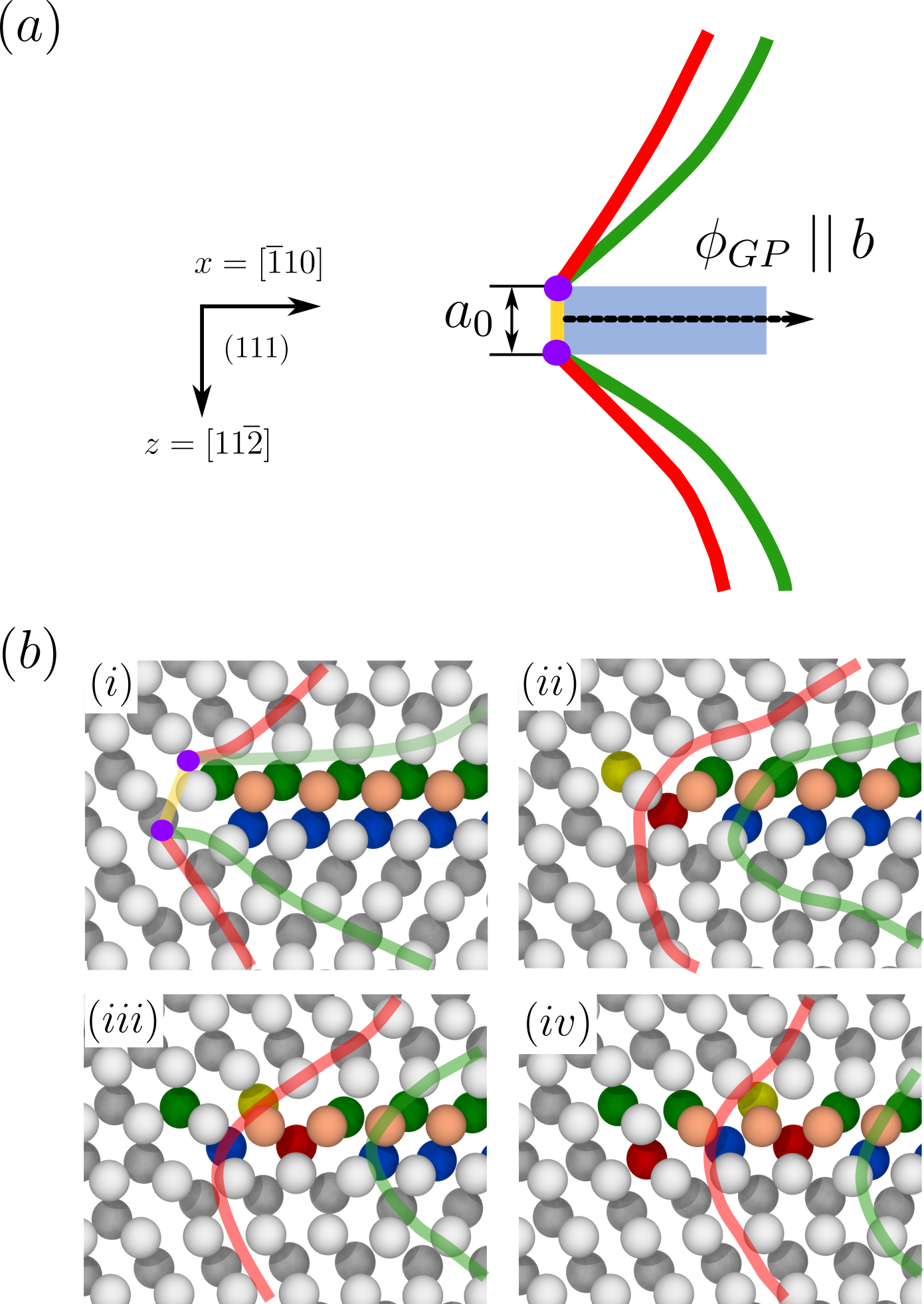} 
\caption{(a) Schematic representation of the initial position of the Shockley partial dislocations with respect to the GP zone in the GP60 orientation. (b) Atomistic representation of the shearing of the GP zone by the Shockley partial dislocations along the (111) plane. Starting from the initial configuration in $(i)$, the progression of the leading and trailing partials through the GP zone is shown from $(ii)$ to $(iv)$. The leading partial is shown as a green line and the trailing partial as a red line. Blue dots stand for Stroh nodes and the full dislocation is shown as a yellow line. Al and Cu atoms that do not move are shown as grey and brown spheres, respectively, while Al and Cu atoms that move during shearing of the GP zone are shown as green and blue spheres, respectively. Atoms with light colors are in the upper plane and with dark color in the lower plane.}
\label{GP0}
\end{figure}

In the GP0 orientation, the first energetic barrier is found at an applied shear strain of 0.12\%, leading to a local minimum in shear stress when the applied shear strain was 0.18\% (Fig. \ref{GP_ss}). The two Shockley partials are re-combined to form a perfect dislocation at the GP zone interface in this metastable situation, as schematically depicted in Fig. \ref{GP0}(a).  The perfect dislocation is limited by two Stroh nodes and the partials remain dissociated within the Al. The perfect dislocation cannot penetrate the GP zone and the shear stress increases with the applied strain until the critical stress of 150 MPa is attained. Then, the perfect dislocation dissociates with the leading partial entering the GP zone and extending  the stacking fault within the GP zone, followed by the propagation of the trailing partial until the GP zone is sheared. The displacement of the Cu atoms in the (111) plane during the shearing of the GP zone by both partials is depicted in Fig. \ref{GP0}(b). The initial atomic configuration, corresponding to  Fig. \ref{GP0}(a), is shown in Fig. \ref{GP0}(b)$(i)$. Starting from this configuration, shearing of the GP by both partials is shown in the atomic arrangements $(ii)$ to $(iv)$. It should be noted that the distance between the partials varied from from 0.7 to 1 nm during the shearing of the GP zone and that the Cu atoms in the slip plane did not move at the same time to the perfect stacking fault positions in the lattice. Moreover, slightly different sequences of atomic displacements to shear the GP zone were observed in different simulations, although the main mechanism did not change. Thus, the analysis of the atomic sequence shows that the major energetic barrier to shear the GP zone in the GP0 orientation is found to split the full dislocation into partials and to introduce the leading partial into the GP zone. This is followed by other smaller barriers associated with the propagation of the leading and trailing partial dislocations along the GP zone. These results are in the agreement with the experimental observations of \cite{Muraishi2002}, who reported that energetic barrier of the  GP zones consists of a number of short range barriers proportional to the diameter of the precipitate along a first order barrier profile. 

\subsection{Activation energy, activation entropy and activation volume}\label{Thermo}

\begin{figure}[!]
\centering
\includegraphics[scale=0.8]{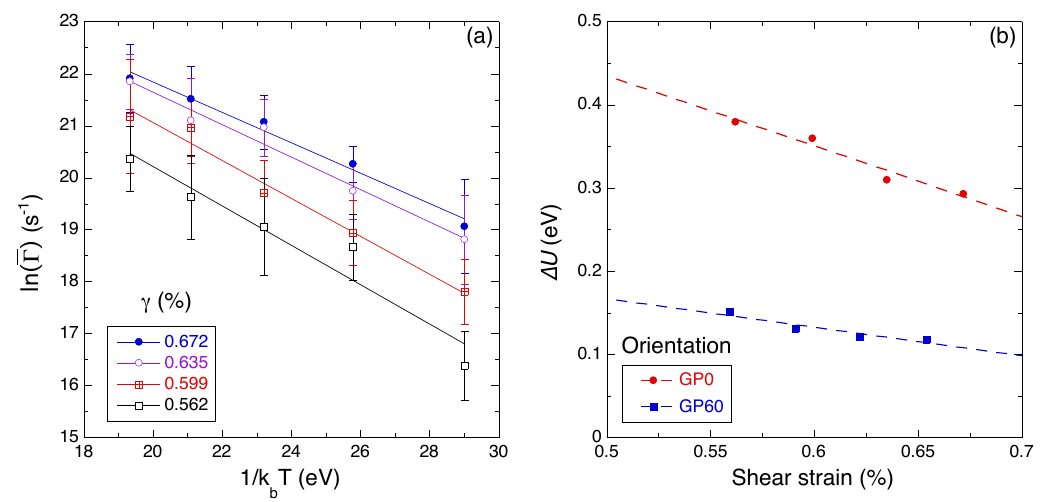} 
\includegraphics[scale=0.8]{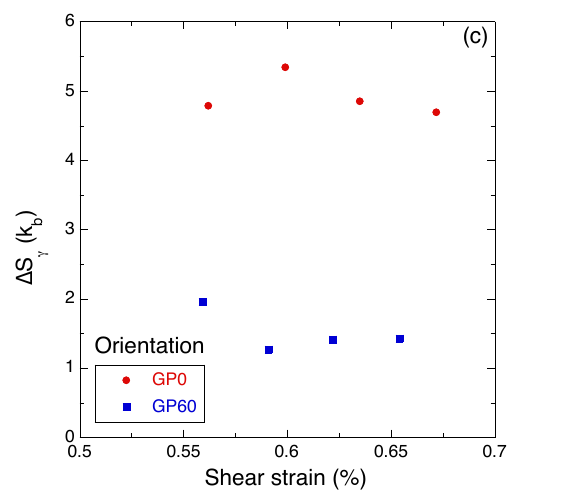} 
\caption{(a) MD results of the natural logarithm of the average rate, $\ln(\overline\Gamma)$ (and standard deviation) at which an edge dislocation shears a GP zone in the GP0 orientation as a function of $1/k_bT$ for different applied shear strains and temperatures. (b) Activation energy, $\Delta U$, as a function of the applied strain $\gamma$ for different dislocation/GP zone orientations. The straight lines stand for the linear extrapolation of the values obtained by MD. (c) Activation entropy, $\Delta S_\gamma$, as a function of the applied strain $\gamma$ for different dislocation/GP zone orientations.}
\label{rate_MD}
\end{figure}

The rate at which the dislocation overcomes the GP zone in the GP0 orientation was determined from MD simulations at different temperatures and applied shear strains. In each case, the initial configuration of the dislocation/GP zone was given by the minimum energy configuration at the given applied shear strain obtained with the MS+TA strategy in the previous section. The average time to  shear the GP zone as a function of the applied shear strain and temperature was determined and its inverse (the average rate $\overline\Gamma$) is plotted as function of $1/k_bT$ in Fig. \ref{rate_MD}(a).  The straight lines in this figure stand for the fitting of the average rates obtained with MD for the same applied strain and different temperatures with eq. \eqref{TST-F}. 
They show that transition state theory is applicable to this phenomenon and that, although multiple barriers have to be overcome to shear the GP zone, the activation free energy can be simplified into a first order saddle point throughout the energy landscape. 

Taking into account eq. \eqref{DeltaF}, eq. \eqref{TST-F} can be expressed as 

\begin{equation}
\ln(\overline\Gamma) =\ln(\nu) + \frac{\Delta S_\gamma}{k_b} -\frac{\Delta U(\gamma)}{k_b T}
\label{TST-U}
\end{equation} 

\noindent and the slope of the straight lines in Fig. \ref{rate_MD}(a) provides the activation energy $\Delta U$ as a function of the applied shear strain. This magnitude is plotted in Fig. \ref{rate_MD}(b) for the interaction of edge dislocations with GP zones in the GP0 and GP60 orientations, the latter obtained by \cite{EMS19} using the same strategy. It is interesting to notice that the energy barrier associated with the GP0 orientation is much higher than that for GP60, although the obstacle strength was slightly lower. These differences are obviously associated with the different shearing mechanisms detailed in section \ref{MS-MSTA}. The higher energy barrier of the GP0 orientation can be attributed to the contribution of the  multiple barriers  along the glide plane after the full dislocation has been dissociated and the leading partial penetrates the precipitate while the energetic barrier in the GP60 orientation only depends on the introduction of the trailing partial into the precipitate.  The energy barrier at zero stress, $\Delta U_0$, as  well as the critical strain equivalent to the CRSS of the obstacle, $\gamma_0$, could be obtained by linear extrapolation of the results in in Fig. \ref{rate_MD}$(b)$ and are reported in Table \ref{Param} for both precipitate orientations.

The activation entropy $\Delta S_\gamma$ could also be obtained from the rates in Fig. \ref{rate_MD}(a) and eq. \eqref{TST-U} assuming that $\nu \approx$ 10$^{10}$ s$^{-1}$ \citep{MR04, sobie2017modal}. The activation entropies are plotted as a function of the applied strain $\gamma$ in Fig. \ref{rate_MD}(c) for the GP0 and GP60 orientations, the latter from the information in \cite{EMS19}. The activation entropy was independent of the applied strain in both orientations but the magnitudes were different. $\Delta S_\gamma$ in the GP60 orientation is in the range 1.3-1.8$k_b$, which is similar to the values reported for the dislocation nucleation \citep{HL10, SR11} and compatible with the entropic contribution associated to the vibrational entropy of solids. The activation entropy in the GP0 orientation  was $\approx 5k_b$, suggesting the presence of anharmonic effects during shearing of the GP zone in this orientation.

\begin{table}
\begin{center}
\caption{Obstacle strength, $\tau_0$, activation energy at zero strain, $\Delta U_0$, critical strain for GP zone shearing, $\gamma_0$, and activation volume, $V_0$, for GP0 and GP60 orientations.} 
\begin{tabular}{|c|c|c|c|c|}
 \hline
 Orientation & $\tau_0$ (MPa) & $\Delta U_0$ (eV) & $\gamma_0$ ($\%$) & $V_0$ \\
  \hline
  GP0  & 150 & 0.86 & 0.85 & 40b$^3$ \\
  GP60 & 186 & 0.45 & 0.80 & 12b$^3$ \\
  \hline
\end{tabular}
\label{Param}
\end{center}
\end{table}

The atomistic simulations can also be used to determine the activation volume, $V_{0}$, which is given by \citep{K75, EMS19}

\begin{equation}\label{V0}
V_{0}= -\left.\frac{\partial \Delta G (\tau)}{\partial \tau}\right|_T
 \simeq -\frac{1}{\mu}\left.\frac{\partial \Delta F (\gamma)}{\partial \gamma}\right|_T = -\frac{1}{\mu'}\frac{\Delta U_0}{\gamma_0}
\end{equation}

\noindent where $\mu'$  is the local shear modulus of the crystal that defines the slope of the $\tau$-$\gamma$ linear relationship between points (i) and (ii) in Fig. \ref{MSTA_GP0}(a). The expressions to the right in eq. \eqref{V0} are valid when $V_0$ is much smaller than the volume of the atomic domain. The corresponding activation volumes for the GP0 and GP60 orientations. They are shown in Table \ref{Param} and reflect the differences in the strength and in the energy barrier between both orientations. They are in good agreement with experimental estimations reported by \cite{BFK61}.

\section{Prediction of flow stress}\label{FlowModel}

The yield strength of a crystal, $\tau_y$, is controlled by the obstacles that act as barriers the dislocation motion. These barriers may be athermal or thermally activated. The first ones ($\tau_{ath}$) depend very weakly on the temperature and are often associated to the elastic interactions between dislocations and to the formation of dislocation networks. In the second ones, the probability to overcome the obstacle is controlled by the thermal fluctuations, leading to a strong dependence of the yield strength ($\tau_{th}$) with the temperature and strain rate \citep{K75}. Both contributions are normally present and have to be taken into account to predict the yield strength according to 

\begin{equation}
\tau_y = \tau_{th} + \tau_{ath}. 
\label{PFlow}
\end{equation} 

The experimental evidence indicates that the athermal contribution to the yield strength can be expressed by the Taylor model \citep{T34} according to

\begin{equation}\label{Taylor}
\tau_{ath}= \alpha \mu b\sqrt{\rho},
\end{equation}

\noindent where $\mu$ = 25 GPa stands for the shear modulus of Al, $b$ = 0.286 nm the Burgers vector \cite{RHL19}, $\rho$ the dislocation density and $\alpha$ = 0.7  is a constant in the range 0.5 - 1 that accounts for the different dislocation interactions \citep{KM03}. Both $\mu$ and $b$ depend weakly on the temperature and, thus, this contribution is practically athermal.

The contribution of the thermal barriers to the yield strength was analyzed by \cite{K75} by means of transition state theory.  The relationship between the free energy barrier, the temperature and the plastic strain rate ($\dot\gamma_p$) was given by  

\begin{equation}
\dot\gamma_p = \dot\gamma_{p_0} \: \exp{\bigg(\frac{-\overline{\Delta G}(\tau)}{k_bT}\bigg)} 
\label{ModelK1}
\end{equation}

\noindent where $\dot\gamma_{p_0}$ the reference plastic strain rate, given by \citep{O34},

\begin{equation}
\dot\gamma_{p0} = \rho b L \nu
\label{ModelK2}
\end{equation}

\noindent where $L$ is the mean free path of the dislocations $\rho$  the dislocation density and $\nu$ the attempt frequency. The term $\overline{\Delta G}$ stands for an effective activation Gibbs free energy that depends on the different types of barriers to the dislocation motion, to be defined below for the interaction of dislocations with GP zones in Al.

If the the volume of the system under investigation is much higher than the activation volume of the process, $\Delta G (\tau) \approx \Delta F ({\gamma})$ \citep{SR11,EMS19} and, hence, eq. \eqref{ModelK1} can be written as

\begin{equation}
\dot\gamma_p = \dot\gamma_{p_0} \: \exp{\bigg(\frac{-\overline{\Delta F}(\gamma)}{k_bT}\bigg)}. 
\label{SrateTST}
\end{equation}

The effect of the applied stress (or strain) on the activation free energy in the case of dislocations that have to overcome obstacles was analyzed in detail in \cite{K75}. For short range obstacles (and neglecting the work carried out by the external force), he proposed the following general phenomenological model for the Helmholtz activation free energy

\begin{equation}
\Delta F(\gamma)= \Delta U_0\left[1-\left(\frac{\gamma}{\gamma_{0}}\right)^{p}\right]^q
\label{Kocks}
\end{equation} 

\noindent where $\Delta U_0$ and $\gamma_0$ were obtained from the atomistic simulations, while $q$ ($1 < q < 2$) and $p$ ($0.5 < p < 1$) are two parameters that control the potential energy profile along the minimum energy path ($q$), and the randomness of the obstacle distribution in the glide plane ($p$). In the case of a periodic array of GP zones, $p$ = 1  \citep{caillard2003thermally}, while the Helmholtz activation  free energy was assumed to be a linear function of $\tau$ in the case of the GP60 orientation ($q$ = 1). 

Combining  eqs. \eqref{ModelK2}, \eqref{Kocks} and \eqref{SrateTST}, and taking into account that there is a linear relationship between the shear stress and strain (Fig. \ref{GP_ss}(a), the thermal flow stress is given by

\begin{equation}
\tau_{th} =  \overline{\tau}_0 \bigg[ 1- \frac{k_b T}{\overline{\Delta U}_0}   \ln \left(\frac{\dot\gamma_{p0}}{\dot\gamma_p}\right)\bigg] 
\label{thermal_flow}
\end{equation}

\noindent where $\tau_0$ stands for the obstacle strength (the stress necessary to overcome the obstacle without thermal assistance) and $\Delta U_0$ denotes the activation energy barrier. 

As indicated above, GP zones grow along three different orientations, corresponding to the three \{100\} faces of the cubic Al FCC lattice, leading to two different orientations between the dislocation and the GP zones, GP0 and GP60. The GP60 orientation is twice more probable than the GP0 \citep{N14, RBP18} and, thus, the effective values of the obstacle strength ($\overline{\tau}_0$) and of the activation energy barrier ($\overline{\Delta U}_0$) in eq. \eqref{thermal_flow} can be determined from those obtained by atomistic simulations for each orientation according to 

\begin{equation}
\overline{\Delta U}_0=\frac{1}{3}\left(2\Delta U_0^{\rm GP60}+\Delta U_0^{\rm GP0}\right)
\label{GeomF}
\end{equation}
\begin{equation}
\overline{\tau}_0=\frac{1}{3}\left(2\tau_0^{\rm GP60}+\tau_0^{\rm GP0}\right).
\label{GeomT}
\end{equation}

\noindent Of course, another type of averaging (e.g. geometric mean) could also be used to determine the effective values of the obstacle strength and of the energy barrier but the differences were negligible in this case.

Thus,  the yield strength of an Al-Cu alloy reinforced with GP zones considering the athermal, eq. \eqref{Taylor}, and thermal, eq. \eqref{thermal_flow},  contributions,  is given by

\begin{equation}
\tau_{y} =  \alpha\mu b\sqrt{\rho} + \overline{\tau}_0 \bigg[ 1- \frac{k_b T}{\overline{\Delta U}_0}   \ln \left(\frac{\dot\gamma_{p0}}{\dot\gamma_p}\right)\bigg].
\label{flow_stress}
\end{equation}

\begin{figure}[t]
\centering
\includegraphics[width=0.45\textwidth]{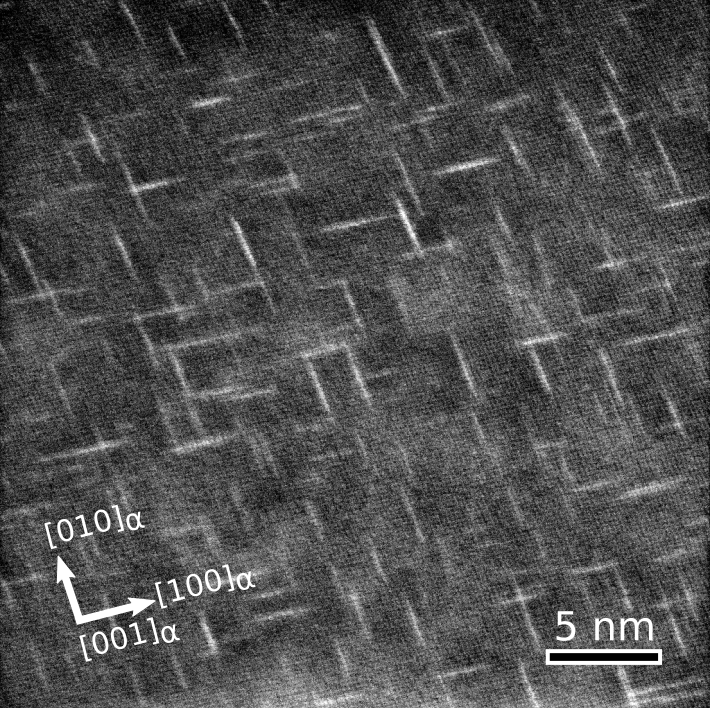} 
\caption{High-angle annular dark-field mode scanning transmission electron microscope micrograph of GP zones in an Al-4 wt. \% Cu alloy aged at ambient temperature for 504 h. The electron beam was close to the [001] orientation of the Al lattice, so GP zones in two different orientations (parallel to the (100) and (010) planes of the Al lattice) are visible.}
\label{TEM}
\end{figure}

\cite{BFK61} measured the influence of the temperature (in the range 4K to 400K)  on $\tau_y$ in single crystals oriented for single slip of a naturally-aged Al - 4 wt. \% Cu alloy. This alloy only contained GP zones but no information about the size and spacing between GP zones and of the initial dislocation density was given. To obtain these data, an Al - 4 wt.\% Cu alloy was manufactured by casting and aged at ambient temperature during 504 hours following the procedure indicated in \cite{RBP18}. Thin foils with a thickness of 110 nm were prepared using a standard lift-out technique and  analysed in a FEI Talos transmission electron microscope (TEM) at 200 kV in a high-angle annular dark-field imaging scanning transmission electron microscope mode in the $\langle100\rangle_{\alpha}$ zone axis. In this orientation, the habit planes of two variants of the GP zones were parallel to the electron beam and a representative micrograph is shown in Fig. \ref{TEM}. The diameter of the GP zones was around 3-4 nm and the distance between them was equivalent to that in the atomistic model ($L$ = 34 nm). Dislocations were observed in a $(1\bar{1}0)$ reflection in a $\langle110\rangle_{\alpha}$ zone axis and the average dislocation density $\rho$ = 2.2 10$^{13}$ m$^{-2}$ was determined from 10 micrographs from the total dislocation length, the foil thickness and the total area analyzed.

Finally, micropillars of 5 x 5 $\mu$m$^2$ cross-section were milled with a focussed ion beam from grains of the Al - 4 wt. \% Cu alloy oriented for single slip following the procedure detailed in \cite{WLA19}. They were tested in compression at 298K and the flow stress was determined from the onset of non linearity in the stress-strain curves. 

\begin{figure}[t]
\centering
\includegraphics[scale=1.0]{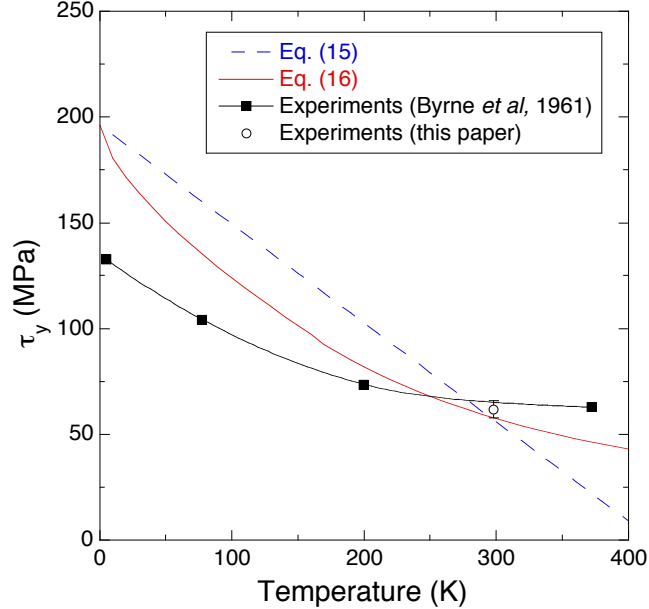} 
\caption{Predictions of the flow stress of an Al alloy containing GP zones as a function of temperature and experimental data from \cite{BFK61}. The dashed blue line stands for a model that accounts for a perfect array of precipitates, eq. \eqref{flow_stress}, while the solid   red line takes into account the random distribution of GP zones, eq. \eqref{LCM}.}
\label{Model}
\end{figure}

The microstructural information obtained from the TEM analysis was introduced in eq. \eqref{flow_stress}, and the prediction of the flow stress as a function of temperature is plotted in Fig. \ref{Model}, assuming a strain rate $\dot\gamma$ = 10$^{-2}$ s$^{-1}$, together with the experimental data from \cite{BFK61} and those obtained from micropillar compression tests at ambient temperature. The linear dependence of the flow stress with temperature was not in agreement with the experimental data  and  one possible reason may be that the thermal contribution in eq. \eqref{flow_stress} assumes a regular distribution of GP zones while they are actually randomly dispersed within the matrix (Fig. \ref{TEM}). The effect of a random precipitate distribution on the thermal contribution to the flow stress was analyzed  by \cite{leyson2012B, LC16}, which modified eq. \eqref{thermal_flow}  to account for this effect, leading to

\begin{equation}
\tau_y =  \alpha \mu b\sqrt{\rho} +
\begin{cases}  
\overline{\tau}_0 \bigg[ 1- \left[ \frac{k_b T}{\overline {\Delta U}_0} \ln \left(\frac{\dot\gamma_{p0}}{\dot\gamma_p} \right) \right]^{\frac{2}{3}}  \bigg]  , & \mbox{if } \:\: \tau_y/ \overline{\tau}_0 \ge 0.55 \\
\\
\overline{\tau}_0 \exp{\bigg[-\frac{1}{0.51}\frac{k_b T}{\overline {\Delta U}_0} \ln \left(\frac{\dot\gamma_{p0}}{\dot\gamma_p}\right)\bigg]}  , & \mbox{if } \:\: \tau_y/\overline{\tau}_0 < 0.55
\end{cases}
\label{LCM}
\end{equation}

The predictions of the flow stress taking into account the random distribution of GP zones  are also plotted in Fig. \ref{Model}, leading to a better agreement with the experimental data. Nevertheless, the model predictions tend to overestimate the strength in the low temperature regime, pointing out to the current limitations of the  methodology. The first one is that the obstacle strength and the energy barrier were only determined for edge dislocations and they may be different from those obtained for screw and mixed dislocations. In addition, there is also an important uncertainty in the linear extrapolation of the activation energies obtain at different values of the applied strain to zero strain. Regardless of these limitations, the overall methodology provides coherent results of the effect of GP zones on the flow stress of Al alloys.

\section{Conclusions} \label{sec6:Conclusion}

A  scale bridging strategy  based in molecular statics and molecular dynamics simulations in combination with transition state theory has been developed to determine the flow stress of Al-Cu alloy containing GP zones as a function of the temperature. The flow stress is assumed to depend on two contributions. The athermal contribution is given by the Taylor model \citep{T34} and only depends on the elastic constants of the Al alloy, the dislocation density and the Burgers vector. The thermal contribution can be calculated following the transition state theory \citep{K75} from the obstacle strength and the free energy barrier. 

Molecular statics simulations in combination with thermal annealing were used to determine the obstacle strength (the stress necessary to overcome the obstacle in the absence of thermal energy) for edge dislocations interacting with GP zones in two different orientations (0 and 60$^\circ$). It was found that the introduction of thermal annealing after each strain increment was critical to avoid the locking of the simulations in local minima which leads to an overestimation of the obstacle strength and to unrealistic interaction mechanisms between the dislocation and the GP zone in both orientations. The rate at which dislocations overcome the GP zone was determined as a a function of the applied strain and temperature from molecular dynamics simulations. This information was used to the determine the thermodynamic quantities that control this process, namely the activation energy, the activation entropy and the activation volume for both dislocation/GP zone orientations.

The predictions of the model were compared with experimental data in the literature as well as new data obtained by means of micropillar compression tests in an Al - 4 wt. \% Cu alloy naturally aged at  ambient temperature to obtain a homogenous distribution of GP zones. They were in good agreement at ambient temperature although overestimated the flow stress at low temperature and the possible sources of the discrepancy were indicated. Overall, the papers presents a coherent  scale bridging  methodology to determine the effect of GP zones in the flow stress of Al alloys by means of atomistic simulations.

\section*{Acknowledgements}
This investigation was supported by the European Research Council under the European UnionÕs Horizon 2020 research and innovation programme (Advanced Grant VIRMETAL, grant agreement No. 669141). The computer resources and the technical assistance provided by the Centro de Supercomputaci\'on y Visualizaci\'on de Madrid (CeSViMa) are gratefully acknowledged. Additionally, the authors thankfully acknowledge the computer resources at Picasso and the technical support provided by University of Malaga and Barcelona Supercomputing Center. Finally, use of the computational resources of the Center for Nanoscale Materials, an Office of Science user facility, supported by the U.S. Department of Energy, Office of Science, Office of Basic Energy Sciences, under Contract No. DE-AC02-06CH11357, is gratefully acknowledged.  BB acknowledges the support from the Spanish Ministry of Education through the Fellowship FPU15/00403. EM used resources provided by the Los Alamos National Laboratory (LANL) Institutional Computing Program. LANL is operated by Triad National Security LLC for the National Nuclear Security Administration of U.S. DOE (contract no. 89233218CNA000001).


\end{document}